# Viscous damping of nanobeam resonators: humidity, thermal noise and the paddling effect

Chao Chen[1,2], Ming Ma[1,2], Jefferson Zhe Liu[3], Quanshui Zheng[1,2] and Zhiping Xu[1,2,*]

[1]Department of Engineering Mechanics, Tsinghua University, Beijing 100084, China

[2]Center for Nano and Micro Mechanics, Tsinghua University, Beijing 100084, China

[3]Department of Mechanical and Aerospace Engineering, Monash University, Clayton, Victoria 3800, Australia

[*]Author to whom correspondence should be addressed. Email address: xuzp@tsinghua.edu.cn

**Abstract**

The nanobeam resonator is the key mechanical component in the nano-electromechanical system. In addition to its high frequency originating from its low dimension, the performance is significantly influenced by the circumstances, especially at nanoscale where a large surface area of the material is exposed. Molecular dynamics simulations and theoretical analysis are used for a quantitative prediction on the damping behavior, such as the critical damping condition and lifetime, of nanobeam resonators that directly maps the fluid-structure properties and interaction information into dynamical behaviors. We show here how the humidity defines the critical damping condition through viscous forces, marking the transition from under-damping to over-damping regime at elevated humidity. Novel phenomena such as the thermal fluctuation and paddling effects are also discussed.

In past decades there are arising interests in fabricating mechanical resonators using nanostructures such as carbon nanotubes (CNTs), zinc oxide nanowires and silicon nanobeams[1-5]. Key advantages of these unique setups include ultra-high frequencies beyond gigahertz, remarkable controllability through electro-mechanical or opto-mechanical coupling, and capability to reach the quantum limit of mechanical systems[6, 7]. Various applications from sensors, actuators to relays are proposed and many of these ideas are already established as building blocks in the nanoelectromechanical systems. However, except for all the aforementioned merits, a critical issue in the way towards practical applications exists as the perturbation from the surrounding environment, which modifies the device behavior significantly at nanoscale[4]. A practical question thus naturally raised for these important applications is that how these devices operate under a humid condition.

In a very recent experiment, carbon nanotubes based resonators are used for ultra-sensitive mass sensing in viscous fluids. It is observed that the resonator loses its fundamental oscillation once immersed in water, while in vacuum a number of resonance modes are still distinct. This result indicates that, in addition to the intrinsic phononic damping, viscous damping also has impacts on perturbing the vibrational motion of carbon nanotubes. Interesting examples showing the importance of the fluid-structure interaction at small scale can also be widely found in living organisms. The over-damped cytoskeleton fibers provide supporting and transport functions in the cell. The unique feature of the dynamics of these fibers is the dominating contribution from thermal fluctuation, as the bending energy is extremely low, even comparable with thermal energy $k_BT$[8]. Thus thermal fluctuation in a fluid environment not only provides viscous damping for the vibration of fibers, but also drives their motion with noticeable amplitudes[9, 10]. The correlation between the thermally induced vibration and environmental temperature is successfully used to estimate the bending rigidity of fibers[9, 11]. Moreover, fluid flow can lead to remarkable deformation or motion of slender fibers, which suggests mechanosensing mechanisms. Some living organisms, *Cupiennius salei* for example, have ultra-vibration-sensitive hairy systems (sensilla) on their body, to detect environmental signals such as air flow in preying.[9] In-depth understanding of these novel systems and phenomena towards bio-inspired applications requires a fundamental investigation on the damping behavior of nanobeam resonators[12].

In this paper, we investigate the damping behavior of a CNT resonator in a water environment with various humidities. Molecular dynamics (MD) simulations are performed to capture the atomistic mechanism, followed by continuum analysis using the Euler-Bernoulli beam theory. The critical damping condition is defined as a function of both properties of the structure and fluid, and their interactions, which is eventually extended to other representative nanobeam resonators.

**Models and Methods**

In our MD simulations, we use the LAMMPS package[13]. A single-walled CNT of length $L_{CNT}$ = 5.5 nm is immersed in a periodic water box with dimensions $L_x$ = 5 nm, $L_y$ = 5 nm in the transverse directions and $L_z$ = 10 nm in the direction in parallel to the CNT. The SPC/E model is used for the intra-molecular interactions (bond stretch and bond angle bending) in the water molecules[14]. This model gives a viscosity $\mu$ = 0.729 mPa s, in close agreement with the experimentally measured value 0.896 mPa s[15]. The SHAKE algorithm is used to constrain the energy terms involving hydrogen atoms, to enable a relatively larger time step $\Delta t$ = 0.5 fs in the simulations[13]. We use the Dreiding force field for the $sp^2$ bonding in the CNT, which is successfully applied to various carbon nanostructures[16, 17]. Van der Waals interactions between the carbon atoms in the CNT and oxygen atoms in the water molecules are described using the Lennard-Jones formula $E_{LJ} = 4\varepsilon[(\sigma/r)^{12} - (\sigma/r)^6]$. This potential function is parameterized to produce the specific contact angle $\theta_{CA}$ = 90° for a water droplet on a graphene sheet, i.e. $\varepsilon$ = 4.0626 meV and $\sigma$ = 0.319 nm[18, 19].

In our MD simulation, the whole system is firstly equilibrated at the ambient condition (temperature $T$ = 300 K, and pressure $P$ = 1 a.t.m.) for 100 ps, with a displacement $d$ = 1.0 nm applied to one end of the CNT to initialize the vibration, while the other end of carbon nanotubes is fixed throughout the simulation.

When it is immersed in a fluid flow, the carbon nanotube is deflected by the viscous damping force. We calculate the relationship between a uniform load $F$ applied transversely on the CNT cantilever beam and its tip deflection $d$. As shown in Figure 1, the tip displacement keeps a linear relationship with $F$ when $d$ is below 1 nm. Beyond 1 nm, the nonlinear effect shows up, and eventually radial buckling occurs when $d$ reaches 2 nm[20]. Thus in our simulations we limit $d$ to be lower than 1 nm, and the linear Euler-Bernoulli beam theory becomes valid. On the other hand, $d$ is also ensured to be larger than the thermal noise to observe an elasticity-driven mechanism, whose amplitude is about 0.15 ~ 0.2 nm, as illustrated by curves in Figure 2.

At time $t$ = 0, the load holding the deflected tip of carbon nanotubes is released, and the CNT starts to be retracted back subject to the elastic restoring force and continues oscillating if not over-damped. In Figure 2, we plot the evolution of tip displacement of the CNT resonator at different nominal humidity $H_N$, which represents the amplitude of first-order vibrational mode of the beam. In this work, $H_N$ is defined as the ratio between the density of water molecules and the equilibrium value of bulk water at 300 K and 1 a.t.m.. The conventional definition of humidity is to quantify the amount of water vapor that exists in the air-water vapor mixed environment. In this paper, $H_N$ is used as an intuitive representation of the population of water molecules around the solid structures. Other constitutes such as the oxygen and nitrogen molecules are neglected and we consider the water in a vapor phase only.

**Results and Discussion**

*Viscous damping*

We perform series of simulations with $H_N$ ranging from 0 to 100 %. It is clearly shown in Figure 2 that the vibrations of CNT beams are significantly damped in humid environments. In vacuum ($H_N = 0$), a distinct under-damping first-order vibration mode of the CNT is observed, with an ultra-high frequency $f_1 = 88.9$ GHz originated from the high Young's modulus $Y$ and small specific-ratio. As $H_N$ increases, at $H_N = 10$ % for example, a significant decay of vibrational amplitude is observed, but the overall behavior in the first several periods is still in the under-damped regime. The phase-shift of the vibration in comparison with the vibration in vacuum is negligible. As $H_N$ further increases to 40 %, the third peak cannot be read out from the noisy background, and the dynamics has a transition from under-damping to over-damping, although the second peak is still noticeable that corresponds to the *paddling effect* as we will discuss later on. Considering all these results and effects we define the condition of critical damping as where the third peak undistinguished in the background noise while second one exists. By this definition the transition from under- to over-damping starts at $H_N \sim 40$ %. Additionally, for all the simulations at $H_N < 40$ %, we notice that the moving tip has a similar speed value $v = 400$ m/s during the first retraction period, indicating an elasticity-driven mechanism and the speed is defined by the bending rigidity of CNTs. However as $H_N$ increases, $v$ continues decreasing to 100 m/s at $H_N = 100$ %, turning into the scenario involving diffusive motion.

*The paddling effects*

Previously we discuss the damping behavior of CNT beam resonators. It is worthy emphasizing that the second peak of vibrational amplitude is observed in all the molecular dynamics simulations without dependence on the humidity, or viscosity. It is noticeable that after the displaced tip of CNT returns to the original (equilibrium) position $x_0$ for the first time, as pointed by the arrow in Figure 2, it always moves further beyond $x_0$, even in an over-damping regime, such as when $H_N$ reaches 100 %. This observation contradicts the picture in a simple damped oscillator model. As the carbon nanotube is moving in the water, it drives the surrounding fluid to flow in the same direction. When the CNT returns to the original position $x_0$ and releases all the elastic potential energy in bending, it does not stop completely. Instead, small amount of momentum is transferred from the fluid back to the CNT, driving the continuous motion. We call this effect as the paddling effect. After the nanobeam loses its elastically driven motion, the tip continues to move due to the thermal fluctuation with amplitude of few Angstroms.

To obtain a deeper insight into the *paddling effect*, here we provide a simple one-dimension model. In this model, as illustrated in the inset of Figure 3, a slider at the position $x$ and with a mass $m_1$ is immersed in a viscous fluid. The slider is also connected with a spring with stiffness $k$, representing the cantilever beam resonator we study in the MD simulations. In this nanoscale flow with a low Reynold number, the viscous drag force can be described accurately by the linear creeping flow equation[21]. Denoting the speed of fluid flow as $v$, we can write the equilibrium equations for both the slider and fluid as:

$$m_1 d^2x/dt^2 + c(dx/dt - v) + kx = 0 \tag{1a}$$

$$m_2 dv/dt - c(dx/dt - v) = 0 \tag{1b}$$

where $m_2$ is the mass of fluid, and $c$ is the drag coefficient. For a cylinder structure, i.e. the carbon nanotube, immersed in the viscous flow, $c = C_d \mu L_{CNT}/2$ where $C_d$ is the viscous damping coefficient. We then have

$$d^2x/dt^2 + 2\xi\omega_0(dx/dt - v) + \omega_0^2 x = 0 \tag{2c}$$

$$m^* dv/dt - 2\xi\omega_0(dx/dt - v) = 0 \tag{2d}$$

where $m^* = m_2/m_1$, $\omega_0 = (k/m_1)^{1/2}$ and $\xi = c/2m_1\omega_0$.

Use the initial displacement amplitude $d$ as a reference for the length scale and $1/\omega_0 = (m_1/k)^{1/2}$ as a reference for the time scale, we obtain the equations in a dimensionless form

$$d^2\underline{x}/d\underline{t}^2 + 2\xi(d\underline{x}/d\underline{t} - \underline{v}) + \underline{x} = 0 \tag{2c}$$

$$m^* d\underline{v}/d\underline{t} - 2\xi(d\underline{x}/d\underline{t} - \underline{v}) = 0 \tag{2d}$$

Here $\underline{x} = x/d$, $\underline{t} = t\omega_0$, $\underline{v} = v/d\omega_0$.

As it is difficult to find analytical solutions for Eq. (2c) and (2d), we present here numerical solutions obtained by using finite difference methods. We plot in Figure 3 the curve of damped vibrational amplitude with initial conditions $x(0) = -1$ and $v(0) = 0$ by setting $m^* = 15$. In this case, the *paddling effect* does influent the damping behavior as it reaches critical value $\xi = 1$, where the slider will always move back over the equilibrium position $x = 0$. Moreover, even in an obviously over-damping regime the paddle effect still exists ($\xi = 1.5$), in consistent with our MD simulations.

*A viscous damping model*

The MD simulations, although providing atomic details of the whole dissipation process, cannot be extend to large length scales and widely spanned fluid and structure types as encountered in the design of micro and nanoelectromechanical systems. To overcome this limitation, we perform theoretical analysis using the Euler-Bernoulli beam theory[14]. The mechanical vibration of an elastic beam can be expressed by the equation

$$D \partial^4 w/\partial x^4 + \rho A \partial^2 w/\partial t^2 + C_d \mu/2 \partial w/\partial t = 0 \quad (3)$$

where $w$ is the deflection of the CNT, $t$ represents time, $x$ is the position along carbon nanotube. $Y$ is the Young's modulus, $I$ is bending moment of inertia of the CNT and $D = YI$ is the bending rigidity. $\rho$ is the mass density and $A$ is the cross-section area. $C$ is the drag coefficient and $\mu$ is the viscosity of the fluid. To determine $C$, we perform MD simulations for water flow around a CNT with the same geometry. For flow in/around nanostructures the Reynolds number Re is usually very low (Re = $\rho u l/\mu < 0.4$, where $\rho$ is the density of water at the ambient condition, and $l$ is characteristic length scale of structure. Here we use $l = D$, the diameter of CNT and find that even at a respectable flow rate ($u_{max} \sim$ 400 m/s) the Reynold number Re = 0.37 is still low.

From molecular dynamics simulations where water flow is induced around carbon nanotubes, we find that at low Reynold number Re < 0.3, the Stokes law $C_d = C/Re$ works well in fitting the simulation results and it yields $C$ = 4.467. For the cantilever beam, one end of the beam ($x$ = 0) is fixed and the other end ($x = L_{CNT}$) is free. By substituting these boundary conditions, i.e. $w(0, t) = \partial w/\partial x(0, t) = \partial w/\partial x(0, t) = \partial^2 w/\partial x^2(L_{CNT}, t) = \partial^3 w/\partial x^3(L_{CNT}, t) = 0$ into Eq. (3), we obtain the eigen-frequencies for vibrational modes n in the absence of damping as

$$\omega_n = \beta_n^2 / L^2 \sqrt{D/\rho A} \quad (4)$$

where $\beta$ is a numeric factor specifically for each vibrational modes. For mode $i$ (= 1, 2, …), $\beta_i$ = 1.875, 4.694, …[22].

In the under-damping regime, the solution of Eq. (3) is still in an oscillational form but the vibrational frequency is shifted[16]. The amplitude $A(t)$ of vibration decays exponentially, i.e. $A(t) = A(0)\exp(-\xi \omega t)$ and the frequency is:

$$\omega_n' = \omega_n \sqrt{1-\xi^2}, 0 < \xi = C\mu L^2 / \left(4\beta_n^2 \sqrt{\rho AD}\right) < 1 \quad (5)$$

Close inspection of our MD simulation results in Figure 2 shows that amplitude $A$ indeed exhibits an exponential decaying. The damping coefficient $\xi$ can thus be fitted. The quality factor $Q$, which is used to quantify the damping effects in the under-damping regime, can be calculated as $Q = 1/(2\xi)$. The results are depicted in Figure 4. As the humidity increases, the quality factor is reduced from the order of $10^3$ in vacuum to 28 at $H_N$ = 2 %, and further under 5 when $H_N$ exceeds 12 %. Due to the paddling effect in our MD simulations, it is hard to define a reasonable quality factor as $Q$ is close to 0.5 ($\xi = 1$), under critical damping conditions.

In the over-damping regime, the oscillation behavior of the beam-resonators is heavily inhibited and one full period of vibration cannot be completed. The frequency becomes

$$\omega_n' = \omega_n\sqrt{\xi^2 - 1}, \ \xi > 1 \tag{6}$$

Figure 2 shows such over-damping vibration at nominal humidities of 60 % and 100 %.

*Lifetime of the resonator*

For the CNT resonator whose vibration is over-damped, its function actually turns into a relay and the key variable to quantify its dynamical behavior is the lifetime of the initialized vibration, or the *returning time* to the equilibrium position $x_0$, instead of the quality factor $Q$. For the humidity $H_N > 10\%$, the vibration energy dissipates within a relatively short time scale of tens of picoseconds. For under-damping vibration, the oscillation amplitude will decay as exponential tendency: $A = A_0\exp(-\xi \omega_n t)$. Here we define the lifetime as $\tau = (\xi \omega_n)^{-1} = 4\rho A/C\mu$, which means the vibrational amplitude decays to 36.8% within the time range $\tau$. For the $\xi$ values as plotted in Figure 4, in most cases with $H_N > 10\%$, $\xi$ is larger than 0.05. So we estimate the lifetime to be shorter than 35.8 ps. According to this reason we plot damping curves in Figure 2 only for the first 40 picoseconds, which can illustrate the whole damping process for most cases, although our total simulation time is much longer.

*Critical damping*

To identify the transition from under-damping to over-damping is crucial to design the micro and nano-resonator. Because of the influences of thermal noise and paddling effect, it's hard to strictly identify the transition from under-damping regime to over-damping directly from our MD simulations. From the Euler-Bernoulli beam theory, by comparing equations (5) and (6), the critical damping, by neglecting the thermal fluctuation and paddling effects as observed in MD simulations, can be defined by $\xi = 1$, i.e.

$$\mu_{c,n} = 4\beta_n^2\sqrt{\rho AD}/CL^2 \tag{7}$$

and for the first-order vibrational mode it is $\mu_{c,1} = 14\sqrt{\rho AD}/CL^2$.

From Eq. (7) we can see that not only the structural properties ($D$, $A$, $\rho$ and $L$) of beams, but also the fluid ($\mu_c$) and the fluid-structure interaction parameters ($C$) play roles in define the critical damping. For a number of types of nanobeams, we predict the behavior of their vibration motion in various fluid environments, as summarized in Figure 5. It can be seen that for the single-walled CNT (diameter $d =$ 1.4 nm as we investigated using MD simulations), a fluidic viscosity leads to over-damping for length $L$ larger than 5 nm. However, at an air environment, it works in an under-damping mechanism. The silicon carbide ($d$ = 20 nm, $L$ = 1 μm) and zinc oxide beam ($d$ = 50 nm, $L$ = 1 μm) resonators are under over-damping and under-damping respectively. For a microtubule ($d$ = 50 nm, $L$ = 5 μm), the bending motion is heavily over-damped due to the relatively low bending rigidity and large aspect ratio. The conclusion from these results can be used to design nano-electromechanical systems in an aqueous or gaseous environment, enabling devices from resonators in the under-damping condition to relays in the over-damping condition. Also they could be inspiring to understand biomechanical systems in close tie to fluid environments, such as aforementioned cytoskelecton networks, the hairy sensory and energy harvesting systems of some insects.

**Conclusion**

In summary, we investigate the viscous damping on nanobeam resonators. The effects of humidity, viscosity and novel phenomena including the thermal noise and paddling effect are discussed, based on results from molecular dynamics simulations and theoretical analysis. The critical damping condition and lifetime of nanobeam resonators are obtained from the analytical models proposed here. The conclusion obtained here provides detailed information for nanoscale fluid-structure interactions and can find applications in both designing nanoelectromechanical systems and understanding related biomechanical systems.

**Acknowledgement**

This work is supported by Tsinghua University through the Key Talent Support Program, and the National Science Foundation of China through Young Scholar Grant 11002079 (ZX), No.10872114, No.10672089 and No. 10832005 (QZ). This work is also supported by Shanghai Supercomputer Center of China.

(Wiley-Interscience, Hoboken, NJ, 2001).

**FIGURES AND FIGURE CAPTIONS**

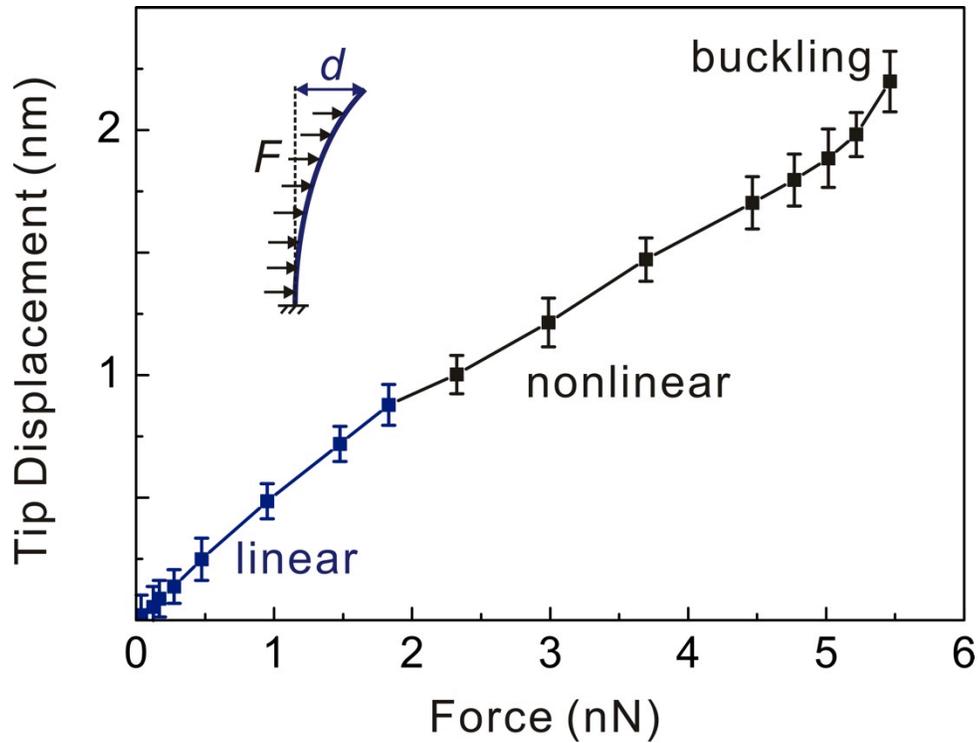

**Figure 1**. The force-displacement relationship for a (10, 10) CNT with a length $L_{CNT}$ = 5.5 nm. The load is distributed uniformly along CNT and deformation is presented by tip displacement $d$. The behavior can be separated into three regimes, linear elastic ($d$ < 1 nm), nonlinear (1 nm < $d$ < 2 nm), and post-buckling regime ($d$ > 2 nm).

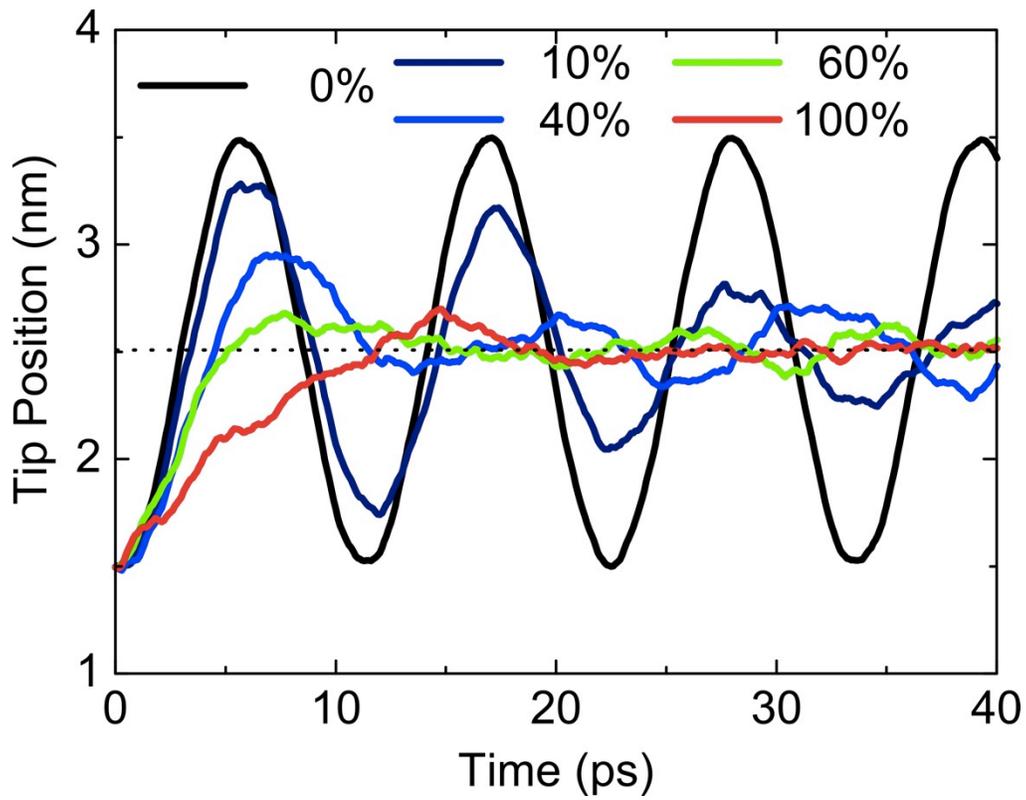

**Figure 2**. The vibrational evolutions of CNT tip positions at different nominal humidity $H_N$ from 0 to 100 %. The difference between under-damping in the vacuum ($H_N = 0$) and over-damping at higher nominal humidities than 40 % is clearly shown.

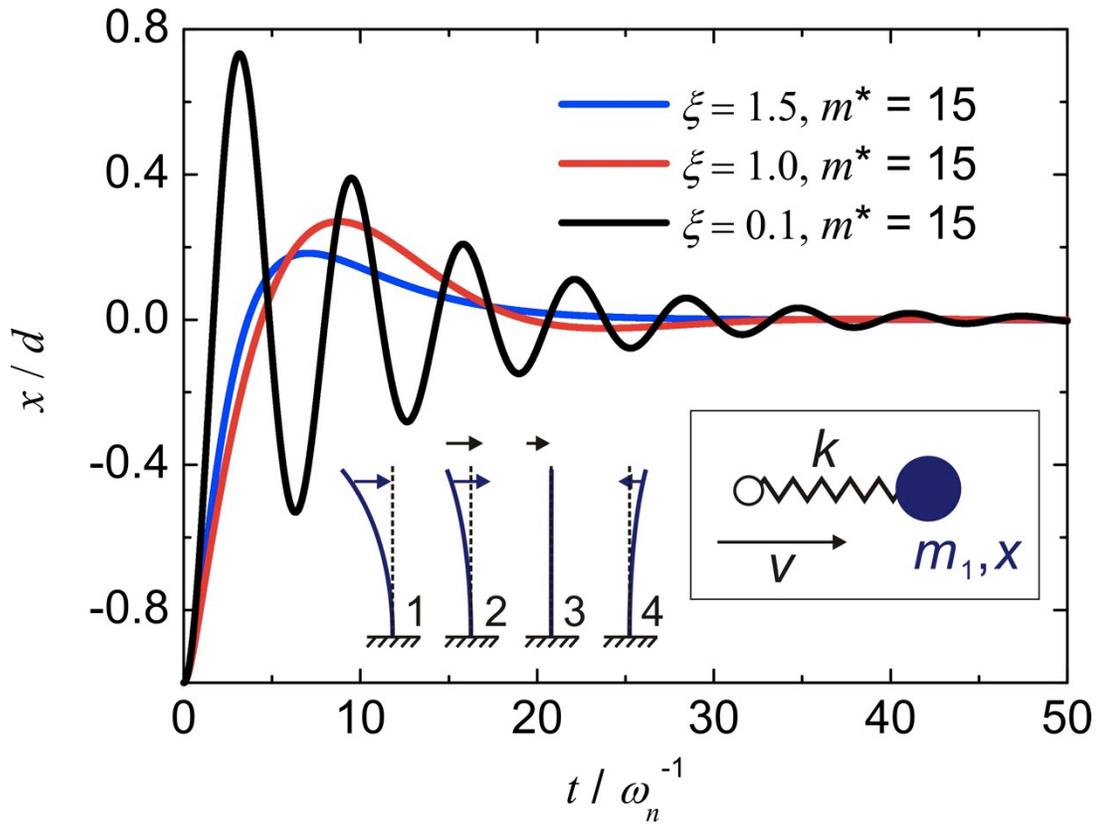

**Figure 3.** The damped vibrational amplitudes of the nanobeam resonators where the paddling effect is included, obtained as solutions of Eqs. (2c) and 2(d). Insets: an illustration of the paddling effect and the analytic model proposed in this work.

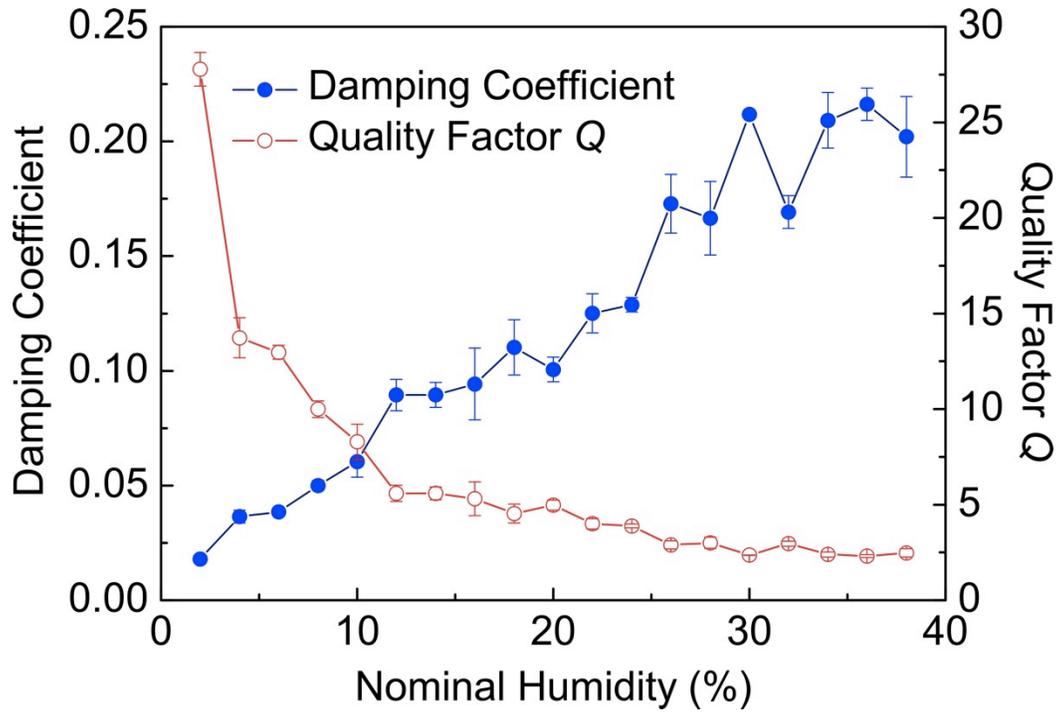

**Figure 4**. Damping coefficients $\xi$ and quality factors $Q$ as obtained from molecular dynamics simulations at different nominal humidity $H_N$ from 0 to 40 %.

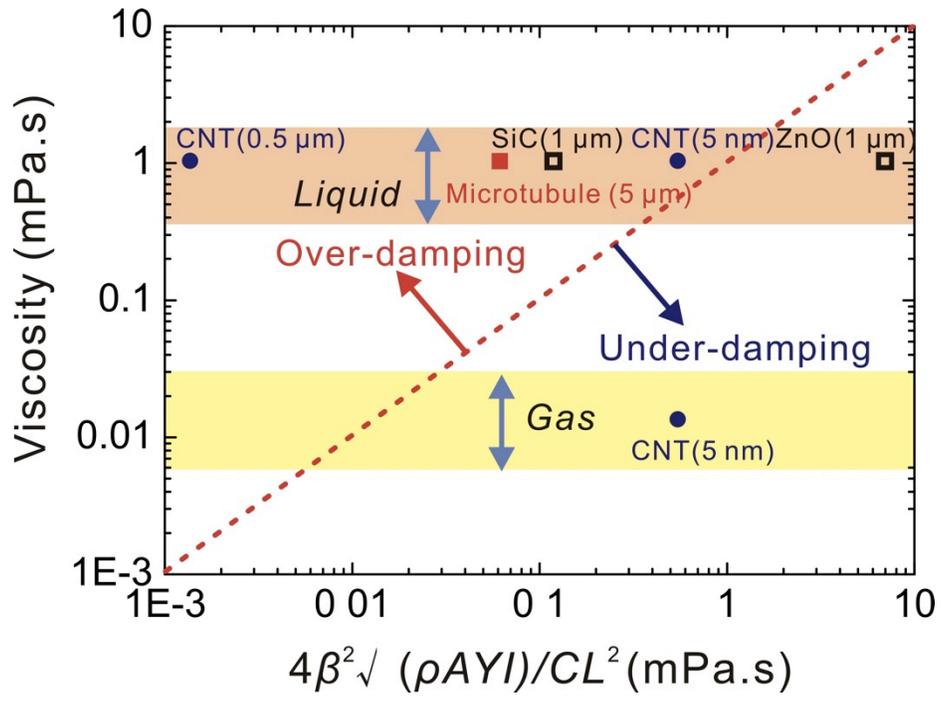

**Figure 5**. Damping behaviors of different types of nano- and micro-fibers, as immersed in various environmental fluids from water to air. The corresponding structural parameters can be found in the text.